\documentclass[conference]{IEEEtran}
\IEEEoverridecommandlockouts
\usepackage{cite}
\usepackage{amsmath,amssymb,amsfonts}
\usepackage{graphicx}
\usepackage{textcomp}
\usepackage{xcolor}
\def\BibTeX{{\rm B\kern-.05em{\sc i\kern-.025em b}\kern-.08em
    T\kern-.1667em\lower.7ex\hbox{E}\kern-.125emX}}
  
\usepackage{amssymb}
\usepackage{gensymb}
\usepackage{graphicx}
\graphicspath{ {images/} }
\usepackage{amsmath}
\usepackage{float}
\usepackage{algorithm}
\usepackage{algpseudocode}
\usepackage{makecell}

\usepackage{geometry}
\usepackage{bm}

\begin{document}

\title{ Rate-Splitting Multiple Access for
Dual-Functional Radar-Communication 
Satellite Systems }

\author{\IEEEauthorblockN{Longfei Yin and Bruno Clerckx}
\IEEEauthorblockA{Department of Electrical and Electronic Engineering, Imperial College London, United Kingdom\\
$\mathrm{Email:}\left \{\mathrm{longfei.yin17, b.clerckx} \right \}\mathrm{@imperial.ac.uk}$
\thanks{
The authors thank the SatnexV Network of Excellence of the European Space Agency for the fruitful discussions, opinions, interpretations, recommendations, and conclusions. Expressed herein are those of the authors and are not necessarily endorsed by the European Space Agency.
}
}
}


\maketitle

\begin{abstract}
In this paper, we consider a multi-antenna dual-functional radar-communication (DFRC) satellite system,
where the satellite has a dual capability to simultaneously communicate with downlink satellite users (SUs) and probe detection signals to a moving target.
To design an appropriate DFRC waveform, we investigate the rate-splitting multiple access (RSMA)-assisted DFRC beamfoming, and 
employ the Cram\'{e}r-Rao bound (CRB) as a radar performance metric, which represents a lower bound on the variance of
unbiased estimators.
The beamforming is optimized to minimize the CRB subject to 
quality of
service (QoS) constraints of SUs and a per-feed transmit power budget.
Satellite communication and 
detecting ground/ sea objects in a bistatic mode are accomplished simultaneously using the DFRC waveform we designed.
Simulation results demonstrate that the proposed RSMA-assisted DFRC beamforming outperforms the conventional space-division multiple access (SDMA) strategy in terms of the communication-sensing trade-off and target estimation performance in a multibeam satellite system.
\end{abstract}

\begin{IEEEkeywords}
Rate-splitting multiple access (RSMA), dual-functional radar-communication (DFRC), multibeam satellite systems, beamforming, bistatic MIMO radar, Cram\'{e}r-Rao bound 
\end{IEEEkeywords}

\section{Introduction}
In 5G and beyond, sharing of the frequency bands between radar sensors and communication systems has
received considerable attention from both industry and academia, and therefore 
motivating the research on dual-functional radar-communication (DFRC) systems.
DFRC techniques 
focus on designing joint systems that can simultaneously perform wireless communication and remote sensing.
Both functionalities are combined via shared use of the spectrum, the hardware platform and a joint signal processing framework \cite{8999605}.
DFRC 
has been considered in several promising terrestrial applications, including autonomous vehicles, human activity monitoring, indoor positioning, etc \cite{8999605,8246850,7060497}.
In \cite{liu2018mu}, a novel framework was proposed for the transmit beamforming of the joint multi-antenna radar-communication (RadCom) system. The precoders are designed to formulate an appropriate desired radar beampattern, while guaranteeing the SINR requirements of the communication users.
\cite{liu2018toward} investigated DFRC waveform optimization approaches
to minimize the multi-user interference (MUI) while formulating a desired radar beampattern.
As a step further, \cite{liu2021cramer} 
proposed a framework for multi-user multi-antenna DFRC beamforming, with emphasis on the optimization of target estimation performance, measured by the Cram\'{e}r-Rao bound (CRB) under both point and extended target scenarios.

In terms of non-terrestrial systems,
as the growing number of communication equipment and various types of radars placed on satellites,
dual-functional waveform design to support simultaneous satellite communication and sensing become necessary to explore.
Rate-splitting multiple access (RSMA), relying on
linearly precoded rate-splitting (RS) at the transmitter and successive interference cancellation (SIC) at the receivers 
has recently emerged as a powerful non-orthogonal transmission and robust interference management strategy for multi-antenna wireless networks.
The benefits achieved by RSMA come from
partially decoding interference and partially treating it as noise, through message splitting and 
have been demonstrated in various multi-antenna scenarios \cite{clerckx2016rate, joudeh2016sum, mao2019rate,  yin2021rate}.
RSMA is also proven to be
promising for multibeam satellite systems whereby there are various practical challenges
including frame-based processing, CSIT uncertainty, per-antenna power constraints, uneven user distribution
and feeder-link interference \cite{yin2020rate, si2021rate}.
Spurred by such advantages, \cite{xu2020rate,xu2021rate} proposed and demonstrated the benefits of a RSMA-assisted DFRC system with the objective to jointly maximize Weighted Sum Rate (WSR) and minimize Mean Square Error (MSE) of beampattern approximation under the per-antenna power constraint. 

Different from aforementioned works, in this paper, we investigate a RSMA-assisted DFRC satellite system which
facilitates the integration of communications and moving target sensing in space networks to make a better use of the
resources, RF spectrum and infrastructure.
Rather than using the MSE of transmit beampattern matching as a radar metric in \cite{xu2020rate,xu2021rate}, explicit optimization of estimation performance at the radar receiver is studied.
The DFRC beamforming optimization is designed to minimize the CRB of target estimation while guaranteeing the quality of
service (QoS) constraints of satellite users (SUs).
To solve the formulated non-convex problem efficiently, we equivalently transform it into  semidefinite programming (SDP), and propose an iterative algorithm based on sequential
convex approximation (SCA) to solve the optimization.
Relying on the Capon method, the parameter estimation of a moving target is evaluated at the radar receiver side. 
Numerical results show that  RSMA is a very promising strategy for the DFRC satellite system to enable a better trade-off between communication and sensing, and achieve better target estimation performance than space-division multiple access (SDMA).

\section{System Model}
\begin{figure}
\centering
\includegraphics[width=0.98 \columnwidth]{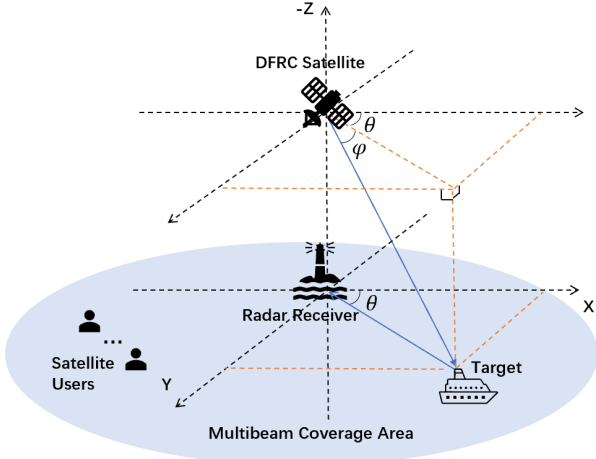}
\caption{Model of a DFRC satellite system.
}
\label{fig:system+}
\end{figure}

We consider a downlink bistatic DFRC satellite system, which simultaneously serves $K$ single-antenna SUs and detects a moving target within the satellite coverage area.
The transmitter is a LEO satellite equipped with $N_{t}$ transmit antennas, while the receiver equipped with $N_{r}$ receive antennas is assumed to be mounted on a buoy or on a balloon.
The model of a DFRC satellite system is illustrated in Fig. \ref{fig:system+}.

When RSMA is adopted,
the messages $W_{1},\cdots, W_{K}$ intended for the SUs indexed by $\mathcal{K} = \left \{ 1, \cdots, K \right \}$ are split into common parts and private parts.
All common parts $ \left \{ W_{c,1}, \cdots, W_{c,K} \right \}$ are jointly encoded into a common stream $s_{c}$, while all private parts $ \left \{ W_{p,1}, \cdots, W_{p,K} \right \}$ are respectively encoded into private streams $s_{1},\cdots,s_{K}$.
Hence, we can denote $\mathbf{s}\left [ l \right ]=\left [ s_{c}\left [ l \right ], s_{1}\left [ l \right ],\cdots, s_{K}\left [ l \right ]\right ]^{T} $ as a $ \left (K+1   \right )\times 1$ vector of unit-power symbol streams, where $l \in \mathcal{L}=\left \{ 1, \cdots, L \right \}$ is the discrete-time index, and the transmit DFRC signal at time index $l$ writes as
\begin{align}
    \mathbf{x}\left [ l \right ] 
    =  \mathbf{P}\mathbf{s}\left [ l \right ] 
    = \mathbf{p}_{c}s_{c}\left [ l \right ]+ 
    \sum _{k \in \mathcal{K}} \mathbf{p}_{k}s_{k}\left [ l \right ] .
    \label{transmit}
\end{align}
$\mathbf{P} = \left [\mathbf{p}_{c},\mathbf{p}_{1},\cdots, \mathbf{p}_{K}  \right ]\in \mathbb{C}^{N_{t} \times \left ( K+1 \right )}$ 
is the dual-functional beamforming matrix to be designed.
If $L$ is sufficiently large, and
the data streams are assumed to be independent from each other, satisfying $\frac{1}{L} \sum_{l=1}^{L}\mathbf{s}\left [ l \right ]  \mathbf{s}\left [ l \right ]^{H}  = \mathbf{I}$.
The covariance of the transmit DFRC signal is given by
\begin{align}
    \mathbf{R}_{X} = 
     \frac{1}{L} \sum_{l=1}^{L}  \mathbf{x}\left [ l \right ]  \mathbf{x}\left [ l \right ]^{H} 
    =\mathbf{P}\mathbf{P}^{H}.
\end{align}
Due to the lack of flexibility in sharing energy resources amongst satellite feeds, we consider a per-feed available power constraint, which is
\begin{align*}
    \mathrm{diag}\left ( \mathbf{P}\mathbf{P}^{H} \right )= \frac{P_{t}\mathbf{1}^{N_{t} \times 1}}{N_{t}},
\end{align*}
where $P_{t}$ represents the total transmit power budget.
By defining $\mathbf{H}=\left [ \mathbf{h}_{1}, \cdots \mathbf{h}_{K} \right ]\in \mathbb{C}^{N_{t}\times K}$ as the communication channel matrix between the satellite and SUs, the received signal at SU-$k$ writes as
\begin{align}
    y_{k}\left [ l \right ]
    &= \mathbf{h}_{k}^{H}\mathbf{x}\left [ l \right ]+n_{k}\left [ l \right ]  \notag
    \\ 
    & = \mathbf{h}_{k}^{H}\mathbf{p}_{c}s_{c}\left [ l \right ]+ 
    \mathbf{h}_{k}^{H}\sum _{k \in \mathcal{K}} \mathbf{p}_{k}s_{k}\left [ l \right ]
    +n_{k}\left [ l \right ],\ \forall k\in \mathcal{K}.
\end{align}
According to the signal propagation characteristics of LEO
satellite communications,
the downlink channel $\mathbf{h}_{k}$
between the satellite and the SU-$k$ can be modeled as in \cite{9165811, 7178538}.
$n_{k}\left [ l \right ]\sim \mathcal{CN}\big ( 0,\sigma _{n,k}^{2} \big )$ represents the additive white Gaussian noise (AWGN) with zero mean.
We assume the noise variance $\sigma _{n,k}^{2}=\sigma _{n}^{2}, \ \forall k \in \mathcal{K}$.
Following the decoding order of RSMA, each SU first decodes the common stream by treating all the other streams as noise.
The SINR of decoding $s_{c}$ at SU-$k$ is expressed by
\begin{align}
     \gamma _{c,k}&= \frac{\left | \mathbf{h}_{k}^{H}\mathbf{p}_{c} \right |^{2}}{\sum_{i \in \mathcal{K}}\left | \mathbf{h}_{k}^{H}\mathbf{p}_{i} \right |^{2} + \sigma _{n}^{2}}, \ \forall k\in \mathcal{K}.
\end{align}
$R_{c,k}= \log_{2}\left ( 1+ \gamma _{c,k}\right )$ is the corresponding achievable rate.
To guarantee that
each SU is capable of decoding the common stream, we define the common
rate as
$R_{c}= \min_{k\in \mathcal{K}}  \left \{ R_{c,k} \right \}= \sum _{k \in \mathcal{K}}C_{k}$, where $C_{k}$ is the rate of the common part of the $k$-th SU's message.
After the common stream is re-encoded, precoded and
subtracted from the received signal through SIC, each SU then decodes its desired private stream.
The SINR of decoding $s_{k}$ at SU-$k$ is given by
\begin{align}
    \gamma _{k}&= \frac{\left | \mathbf{h}_{k}^{H}\mathbf{p}_{k} \right |^{2}}{\sum_{i \in \mathcal{K},i \neq k}\left | \mathbf{h}_{k}^{H}\mathbf{p}_{i} \right |^{2} + \sigma _{n}^{2}},\ \forall k\in \mathcal{K}.
\end{align}
The corresponding rate is $R_{k}= \log_{2}\left ( 1+ \gamma _{k}\right )$, and the total achievable rate of SU-$k$ writes as $R_{k,\mathrm{tot}}= C_{k}+ R_{k}, \forall k \in \mathcal{K}$.

Since the transmitted DFRC signal is used to sense the target as well, the $N_{r} \times 1$ reflected echo signal at the receiver writes as
\begin{align}
   \mathbf{z}\left [ l \right ] = \alpha  e^{j 2\pi \mathcal{F}_{D} lT}{\mathbf{b}} \left ( \theta_{b} ,\phi_{b}\right )
 {\mathbf{a}}^{H} \left ( \theta, \phi \right )\mathbf{x}\left [ l \right ] + \mathbf{m}\left [ l \right ],
 \label{receive_sig}
\end{align}
where 
$\alpha$ stands for the complex
reflection coefficient
which is related to the radar cross-section (RCS) of the target.
$\mathcal{F}_{D}$ denotes the Doppler frequency, and $T$ denotes the symbol period.
$\mathbf{a} \left ( \theta ,\phi\right ) \in \mathbb{C}^{N_{t}\times 1}$ and $\mathbf{b} \left ( \theta_{b} ,\phi_{b}\right )  \in \mathbb{C}^{N_{r}\times 1}$
are the transmit and receive steering vector.
$\left ( \theta ,\phi\right ) $ and $\left ( \theta_{b} ,\phi_{b}\right ) $ respectively  denote the azimuth and elevation angles of the target with respect to the satellite and the receiver.
Since we assume the receiver is mounted on a buoy or a balloon locating exactly below the satellite, and its height can be neglected compared with the satellite height, 
(\ref{receive_sig}) is simplified as
\begin{align}
   \mathbf{z}\left [ l \right ] = \alpha  e^{j 2\pi \mathcal{F}_{D} lT} {\mathbf{b}} \left ( \theta \right )
 {\mathbf{a}}^{H} \left ( \theta, \phi \right )\mathbf{x}\left [ l \right ] + \mathbf{m}\left [ l \right ],
\end{align}
where $\mathbf{m}\left [ l \right ]$ is the AWGN vector following $\mathbf{m}\left [ l \right ]\sim \mathcal{CN}\left ( \mathbf{0}_{N_{r}} ,\sigma _{m}^{2}\mathbf{I}_{N_{r}}\right )$, with $\sigma _{m}^{2}$ denoting the variance of each entry.
Specifically, the transmit and receive steering vector are defined as follows
\begin{align}
    &\mathbf{a}\small ( \theta, \phi \small ) = e^{\left ( j \frac{2 \pi }{\lambda } \left [ \bar{\mathbf{r}}_{1} , \cdots ,  \bar{\mathbf{r}}_{N_{t}}\right ]^{T} \left [ \cos \theta \cos \phi, \sin \theta \cos \phi, \sin \phi \right ]^{T}\right )},
    \\
    &\mathbf{b}\small ( \theta \small ) = e^{\left ( - j \frac{2 \pi }{\lambda } \left [ \mathbf{r}_{1} , \cdots ,  \mathbf{r}_{N_{r}}\right ]^{T} \left [ \cos \theta , \sin \theta , 0 \right ]^{T}\right )}.
\end{align}
The matrices $\left [ \bar{\mathbf{r}}_{1} , \cdots ,  \bar{\mathbf{r}}_{N_{t}}\right ] \in \mathcal{R}^{3 \times N_{t} }$ and $\left [ \mathbf{r}_{1} , \cdots ,  \mathbf{r}_{N_{r}}\right ] \in \mathcal{R}^{3 \times N_{r} }$ have columns representing the Cartesian coordinates of the transmit and receive array
elements, respectively.

\section{DFRC Waveform Optimization}
To design the satellite DFRC beamforming matrix, we employ the Cram\'{e}r-Rao bound (CRB) 
as a radar performance metric, which represents 
a lower bound on the variance of unbiased estimators, and employ per-SU achievable rate as a communication performance metric to ensure the quality of-service (QoS).

The CRB matrix is calculated with respect to the target angle parameters $\mbox{\boldmath$\xi$} = [\theta, \phi]^{T}$, and can be written as $\mathbf{C} = \mathbf{F}^{-1}$,
where $\mathbf{F} $ is the Fisher information matrix (FIM) expressed by
\begin{align}
   \mathbf{F} =
   \begin{bmatrix}
F_{\theta \theta} & F_{\theta \phi}\\ 
F_{\theta \phi}^{T} & F_{\phi \phi}
\end{bmatrix}.
\label{FIM}
\end{align}
From \cite{1703855, kay1993fundamentals}, by defining $\mbox{\boldmath$\mu$} \left [ l \right ] = \mathbf{z}\left [ l \right ] - \mathbf{m}\left [ l \right ]$, we have
\begin{align}
    \left [\mathbf{F}  \right ]_{i,j} = \frac{2}{\sigma _{m}^{2}} \mathrm{Re}\Big \{ \sum_{l=1}^{L} \frac{\partial \mbox{\boldmath$\mu$} \left [ l \right ]^{H}}{\partial  \xi_{i}}   \frac{\partial \mbox{\boldmath$\mu$}\left [ l \right ] }{\partial  \xi_{j}}  \Big \}, \ i,j \in \left \{ 1,2 \right \}.
\end{align}
The elements of the FIM are derived as follows
\begin{align}
F_{\theta,\theta}  &
= \frac{2 \left | \alpha \right |^{2} L}{\sigma _{m}^{2}} \mathrm{Re}\Big \{ \mathrm{tr}\Big ( \frac{\partial\mathbf{A}\left ( \theta,\phi \right )}{\partial\theta }  \mathbf{R}_{X}  \frac{\partial\mathbf{A}\left ( \theta,\phi \right )}{\partial\theta }^{H} \Big )   \Big \},
\\
 F_{\theta,\phi} & 
= \frac{2 \left | \alpha \right |^{2} L}{\sigma _{m}^{2}} \mathrm{Re}\Big \{ \mathrm{tr}\Big ( \frac{\partial\mathbf{A}\left ( \theta,\phi \right )}{\partial\phi }  \mathbf{R}_{X}  \frac{\partial\mathbf{A}\left ( \theta,\phi \right )}{\partial\theta }^{H}  \Big )   \Big \},
\\
F_{\phi,\phi} & 
= \frac{2 \left | \alpha \right |^{2} L}{\sigma _{m}^{2}} \mathrm{Re}\Big\{ \mathrm{tr}\Big ( \frac{\partial\mathbf{A}\left ( \theta,\phi \right )}{\partial\phi }  \mathbf{R}_{X}  \frac{\partial\mathbf{A}\left ( \theta,\phi \right )}{\partial\phi }^{H}  \Big )   \Big \},
\end{align}
where we define $\mathbf{A}\left ( \theta,\phi \right ) = \mathbf{b}\left ( \theta \right ) \mathbf{a}^{H} \left ( \theta,\phi \right )$, and
the partial derivatives are given by
\begin{align}
    \frac{\partial\mathbf{A}\left ( \theta,\phi \right )}{\partial\theta } & =  \frac{\partial\mathbf{b}\left ( \theta \right )}{\partial\theta } \mathbf{a}^{H} \left ( \theta,\phi \right ) + \mathbf{b}\left ( \theta \right ) \frac{\partial\mathbf{a} \left ( \theta,\phi \right )}{\partial\theta }  ^{H} ,
    \\
    \frac{\partial\mathbf{A}\left ( \theta,\phi \right )}{\partial\phi } &=   \mathbf{b}\left ( \theta \right ) \frac{\partial\mathbf{a} \left ( \theta,\phi \right )}{\partial\phi }  ^{H} .
\end{align}

Next, we consider the DFRC waveform, or more precisely, the DFRC beamforming optimization.
The problem of minimizing the trace of the CRB matrix subject to the QoS and transmit power budget constraints is formulated as
\begin{subequations}
\begin{align}
    &\min_{\mathbf{P},\mathbf{c}} \mathrm{tr}\left ( \mathbf{C} \right ) \\
    s.t. \quad 
    &\mathrm{diag}\left ( \mathbf{P}\mathbf{P}^{H} \right )= \frac{P_{t}\mathbf{1}^{N_{t} \times 1}}{N_{t}} 
    \label{per-feed}
    \\
    & R_{c,k} \geq \sum_{i=1}^{K} C_{i} , \ \forall k \in \mathcal{K} 
    \label{common_1}
    \\
    & C_{k} \geq 0, \ \forall k \in \mathcal{K} 
    \label{Cgeq}
    \\
    & R_{k} + C_{k} \geq R_{\mathrm{th}}, \ \forall k \in \mathcal{K} 
    \label{rth}
\end{align} \label{original_p}
\end{subequations}
where $\mathbf{c} = \left [ C_{1}, \cdots, C_{K}  \right ]^{T}$ is vector of common rate portions.
$R_{\mathrm{th}}$ represents the required rate for each user.
(\ref{per-feed}) is the per-feed available power constraint, 
(\ref{common_1}) ensures that the common stream can be successfully decoded by all SUs, and (\ref{Cgeq}) guarantees the non-negativity of all common rate portions.
(\ref{rth}) is the communication QoS constraint.
By letting $\mathbf{P}_{c} = \mathbf{p}_{c}\mathbf{p}_{c}^{H}, \mathbf{P}_{k} = \mathbf{p}_{k}\mathbf{p}_{k}^{H}, \mathbf{H}_{k} = \mathbf{H}_{k}\mathbf{H}_{k}^{H}$,
the original problem (\ref{original_p}) can be equivalently transformed into a semi-definite programming (SDP) problem given by
\begin{subequations}
\begin{align}
    &\min_{\left \{ t_{1}, t_{2} \right \}, \mathbf{P}_{c},\left \{ \mathbf{P}_{k}  \right \}_{k=1}^{K} ,\mathbf{c},\mathbf{r}}  \sum_{i=1}^{2} t_{i} \\
    s.t. \quad 
    & \begin{bmatrix}
\mathbf{F} & \mathbf{e}_{i}\\ 
\mathbf{e}_{i}^{T} & t_{i}
\end{bmatrix} \succeq 0,\ i \in \left \{ 1,2 \right \}
\label{sdp_con1}
\\
&\mathrm{diag}\Big ( \mathbf{P}_{c} + \sum_{k=1}^{K}\mathbf{P}_{k} \Big )= \frac{P_{t}\mathbf{1}^{N_{t} \times 1}}{N_{t}} 
\label{sdp_con2}
\\
& \mathbf{P}_{c} \succeq 0,\ \mathbf{P}_{k} \succeq 0,\ \forall k \in \mathcal{K}
\label{sdp_con3}
\\
& \mathrm{rank}\left ( \mathbf{P}_{c} \right ) = 1, \ \mathrm{rank}\left ( \mathbf{P}_{k} \right ) = 1,\ \forall k \in \mathcal{K}
\label{rankone}
\\
& \log_{2}\Big( 1 + \frac{\mathrm{tr}\left ( \mathbf{H}_{k} \mathbf{P}_{c}\right )   }{\sum _{j \in \mathcal{K} }\mathrm{tr}\left ( \mathbf{H}_{k} \mathbf{P}_{j}\right )+ \sigma _{n}^{2}   } \Big ) \geq \sum_{i=1}^{K} C_{i} , 
\label{common_sdp}
\\
& C_{k} \geq 0, \ \forall k \in \mathcal{K} 
\label{sdp_con4}
    \\
& \log_{2}\Big ( 1 + \frac{\mathrm{tr}\left ( \mathbf{H}_{k} \mathbf{P}_{k}\right )   }{\sum _{j \in \mathcal{K} , j \neq k}\mathrm{tr}\left ( \mathbf{H}_{k} \mathbf{P}_{j}\right )+ \sigma _{n}^{2}   } \Big )  \geq r_{k}, 
\label{rth_sdp}
\\
&
C_{k}+ r_{k} \geq R_{\mathrm{th}}, \ \forall k \in \mathcal{K} 
\label{rthrth_sdp}
\end{align} \label{sdp_p}
\end{subequations}
where $\left \{ t_{1}, t_{2} \right \}$, $\mathbf{r}=\left [ r_{1}, \cdots, r_{K} \right ]^{T}$ are auxiliary variables, $\mathbf{e}_{i}$ denotes the $i$-th column of an identity matrix.
The covariance matrix is expressed by $\mathbf{R}_{X} = \mathbf{P}\mathbf{P}^{H} = \mathbf{P}_{c} +  \sum_{k=1}^{K} \mathbf{P}_{k}$.
To deal with the non-convexity of rate constraints (\ref{common_sdp}) and (\ref{rth_sdp}),
we first rewrite them by introducing slack variables $\left \{ \eta _{c,k} \right \}_{k=1}^{K}, \left \{ \beta _{c,k} \right \}_{k=1}^{K}, \left \{ \eta _{k} \right \}_{k=1}^{K}, \left \{ \beta _{k} \right \}_{k=1}^{K}$
\begin{align}
    &\eta _{c,k}- \beta _{c,k} \geq \sum_{i=1}^{K} C_{i} \log 2, \ \forall k \in  \mathcal{K},
    \label{common_sdp_expa}
    \\
    &e^{\eta _{c,k}}\leq \mathrm{tr}\left ( \mathbf{H}_{k} \mathbf{P}_{c}\right ) + \sum _{j \in \mathcal{K} }\mathrm{tr}\left ( \mathbf{H}_{k} \mathbf{P}_{j}\right )+ \sigma _{n}^{2} , \ \forall k \in  \mathcal{K},
    \label{common_sdp_expb}
    \\
    &e^{\beta _{c,k}}\geq \sum _{j \in \mathcal{K} }\mathrm{tr}\left ( \mathbf{H}_{k} \mathbf{P}_{j}\right )+ \sigma _{n}^{2} , \ \forall k \in  \mathcal{K},
    \label{common_sdp_expc}
    \\
    &\eta _{k}- \beta _{k} \geq r_{k} \log 2, \ \forall k \in  \mathcal{K},
    \label{rth_sdp_expa}
    \\
    &e^{\eta _{k}}\leq  \sum _{j \in \mathcal{K} }\mathrm{tr}\left ( \mathbf{H}_{k} \mathbf{P}_{j}\right )+ \sigma _{n}^{2} , \ \forall k \in  \mathcal{K},
    \label{rth_sdp_expb}
    \\
    &e^{\beta _{k}}\geq \sum _{j \in \mathcal{K} , j \neq k}\mathrm{tr}\left ( \mathbf{H}_{k} \mathbf{P}_{j}\right )+ \sigma _{n}^{2} , \ \forall k \in  \mathcal{K}.
    \label{rth_sdp_expc}
\end{align}
Note that (\ref{common_sdp_expc}) and (\ref{rth_sdp_expc}) are still nonconvex with convex left hand sides (LHSs) which can be approximated
by the first-order Taylor approximation given as follows
\begin{align}
     & \sum _{j \in \mathcal{K} }\mathrm{tr}\left ( \mathbf{H}_{k} \mathbf{P}_{j}\right )+ \sigma _{n}^{2} \leq e^{\beta _{c,k}^{\left [ n \right ]}}\big (\beta _{c,k} - \beta _{c,k}^{\left [ n \right ]} +1 \big ), \ \forall k \in  \mathcal{K},
     \label{common_sdp_expc_approx}
     \\
     & \sum _{j \in \mathcal{K},j \neq k }\mathrm{tr}\left ( \mathbf{H}_{k} \mathbf{P}_{j}\right )+ \sigma _{n}^{2} \leq e^{\beta _{k}^{\left [ n \right ]}}\big (\beta _{k} - \beta _{k}^{\left [ n \right ]} +1 \big ), \ \forall k \in  \mathcal{K},
     \label{rth_sdp_expc_approx}
\end{align}
where $n$ represents the $n$-th SCA iteration. 
(\ref{common_sdp_expb}) and (\ref{rth_sdp_expb}) belong to generalized nonlinear convex program, which leads to high computational
complexity. Aiming at more efficient implementation, we introduce 
$\left \{ \tau _{c,k} \right \}_{k=1}^{K},  \left \{ \tau _{k} \right \}_{k=1}^{K}$, and rewrite (\ref{common_sdp_expb}) and (\ref{rth_sdp_expb}) as
\begin{align}
    &\tau _{c,k}  \leq \mathrm{tr}\left ( \mathbf{H}_{k} \mathbf{P}_{c}\right ) + \sum _{j \in \mathcal{K} }\mathrm{tr}\left ( \mathbf{H}_{k} \mathbf{P}_{j}\right )+ \sigma _{n}^{2} , \ \forall k \in  \mathcal{K},
    \label{common_sdp_expb_expa}
    \\
    &\tau _{c,k} \log \left (\tau _{c,k}  \right ) \geq \tau _{c,k} \eta _{c,k}, \ \forall k \in  \mathcal{K},
    \label{common_sdp_expb_expb}
    \\
    &\tau _{k}  \leq   \sum _{j \in \mathcal{K} }\mathrm{tr}\left ( \mathbf{H}_{k} \mathbf{P}_{j}\right )+ \sigma _{n}^{2} , \ \forall k \in  \mathcal{K},
    \label{rth_sdp_expb_expa}
    \\
    &\tau _{k} \log \left (\tau _{k}  \right ) \geq \tau _{k} \eta _{k} , \ \forall k \in  \mathcal{K}.
    \label{rth_sdp_expb_expb}
\end{align}
The LHSs of (\ref{common_sdp_expb_expb}) and (\ref{rth_sdp_expb_expb}) are convex, so we compute the first-order Taylor series approximations, which are respectively
\begin{align}
    &\tau _{c,k}^{\left [ n \right ]} \log \big ( \tau _{c,k}^{\left [ n \right ]} \big ) + \big ( \tau _{c,k}- \tau _{c,k}^{\left [ n \right ]} \big )\big [ \log \big ( \tau _{c,k}^{\left [ n \right ]} \big ) +1 \big ]\geq \tau _{c,k} \eta _{c,k},
    \\
    &\tau _{k}^{\left [ n \right ]} \log \big ( \tau _{k}^{\left [ n \right ]} \big ) + \big ( \tau _{k}- \tau _{k}^{\left [ n \right ]} \big )\big [ \log \big ( \tau _{k}^{\left [ n \right ]} \big ) +1 \big ]\geq \tau _{k} \eta _{k}.
\end{align}
The equivalent second-order cone (SOC) forms are
\begin{align}
    &\Big \| \Big [ \tau_{c,k} + \eta_{c,k}-\big (\log \big (  \tau_{c,k}^{\left [ n \right ]}  \big )+1  \big ) , \ 2\sqrt{ \tau_{c,k}^{\left [ n \right ]} }    \Big ]  \Big \|_{2} \notag
    \\
    &\leq \tau_{c,k} - \eta_{c,k} + \log \big (  \tau_{c,k}^{\left [ n \right ]}  \big )+1,  \ \forall k \in  \mathcal{K},
    \label{common_sdp_expb_expb_approx}
    \\
    &\Big \| \Big [ \tau_{k} + \eta_{k}-\big (\log \big (  \tau_{k}^{\left [ n \right ]}  \big )+1  \big ) , \ 2\sqrt{ \tau_{k}^{\left [ n \right ]} }    \Big ]  \Big \|_{2}
    \notag
    \\
    &\leq \tau_{k} - \eta_{k} + \log \big (  \tau_{k}^{\left [ n \right ]}  \big )+1, \ \forall k \in  \mathcal{K}.
    \label{rth_sdp_expb_expb_approx}
\end{align}
For the nonconvex (\ref{rankone}), we can build an iterative penalty function to insert these rank-one constraints into the objective function.
By defining  $\mathbf{v}_{c,\mathrm{max}}^{\left [ n \right ]}$ as the the normalized eigenvector corresponding to the maximum eigenvalue $\lambda _{\mathrm{max}}\big ( \mathbf{P}_{c}^{\left [ n \right ]} \big )$, and $ \big \{ \mathbf{v}_{k,\mathrm{max}}^{\left [ n \right ]} \big \}_{k=1}^{K}$ as the the normalized eigenvector corresponding to $\big \{  \lambda _{\mathrm{max}}\big ( \mathbf{P}_{k}^{\left [ n \right ]} \big ) \big \}^{K}_{k=1}$,
the above problem (\ref{sdp_p}) can be reformulated by
\begin{align}
    &\min_{\left \{ t_{1}, t_{2} \right \},  \mathbf{P}_{c},  \left \{ \mathbf{P}_{k}  \right \}_{k=1}^{K},\mathbf{c},\mathbf{r},\eta,\beta,\tau}  \sum_{i=1}^{2} t_{i} + \mathrm{PF}  \label{obj}
    \\
    s.t. \   &(\ref{sdp_con1})-(\ref{sdp_con3}),(\ref{sdp_con4}),(\ref{rthrth_sdp}) \notag
    \\
    &
    (\ref{common_sdp_expa}),(\ref{rth_sdp_expa}),
    (\ref{common_sdp_expc_approx}), (\ref{rth_sdp_expc_approx}), 
    (\ref{common_sdp_expb_expa}) \notag
    \\ 
    &(\ref{rth_sdp_expb_expa}),
    (\ref{common_sdp_expb_expb_approx}),(\ref{rth_sdp_expb_expb_approx})
    \notag
\end{align}
where $\eta,\beta,\tau$
are defined as the sets of introduced slack variables.
The iterative penalty function is expressed by
\begin{align}
    \mathrm{PF} = \lambda \Big ( \Big [ \mathrm{tr} \left ( \mathbf{P}_{c} \right ) - \big ( \mathbf{v}_{c,\mathrm{max}}^{\left [ n \right ]} \big )^{H} \mathbf{P}_{c} \mathbf{v}_{c,\mathrm{max}}^{\left [ n \right ]}\Big ] \notag
    \\
    + \sum_{k=1}^{K} \Big [ \mathrm{tr} \left ( \mathbf{P}_{k} \right ) - \big ( \mathbf{v}_{k,\mathrm{max}}^{\left [ n \right ]} \big )^{H} \mathbf{P}_{k} \mathbf{v}_{k,\mathrm{max}}^{\left [ n \right ]}\Big ] \Big ).
\end{align}
$\lambda$ is a penalty factor which is selected properly during simulation to ensure $\mathrm{PF}$ as small as possible.
Problem (\ref{obj}) is convex and can be effectively solved by the CVX toolbox. The results
obtained from the $n$-th iteration are treated as constants while solving (\ref{obj}).
We summarize the procedure of this
DFRC waveform design in Algorithm \ref{algorithm1}. 
Initialization can be the solution of a max-min fair rate problem in a communication-only scenario.
$\varepsilon$ is the
tolerance value. 
The convergence of Algorithm \ref{algorithm1} is guaranteed since the solution of Problem (\ref{obj}) at
iteration-$n$ is a feasible solution of the problem at iteration-$n+1$. 
Finally, eigenvalue decomposition (EVD) can be
used to calculate the optimized beamforming vectors, and the optimized CRB is obtained accordingly.
Note that Problem (\ref{obj}) involves only SOC and linear matrix inequality (LMI) constraints, it can be solved by using interior-point methods with the worst-case computational complexity $\mathcal{O}\big ( \log\small ( e^{-1} \small ) \small [N_{t}^{2}\left ( K+1 \right )  \small ]^{3.5}\big )$ \cite{mao2020max, mao2019rate, ye2011interior}.

\begin{algorithm}
\caption{DFRC Beamforming Optimization}\label{algorithm1}
\begin{algorithmic}
\State \textbf{Initialize}: $n\leftarrow 0,\mathbf{P}_{c}^{\left [ n \right ]},\big \{ \mathbf{P}_{k}^{\left [ n \right ]} \big \}_{k=1}^{K},\eta^{\left [ n \right ]},\beta^{\left [ n \right ]},\tau^{\left [ n \right ]}$;
\Repeat
\State Solve (\ref{obj}) at 
$\mathbf{P}_{c}^{\left [ n \right ]},\big \{ \mathbf{P}_{k}^{\left [ n \right ]} \big \}_{k=1}^{K},\eta^{\left [ n \right ]},\beta^{\left [ n \right ]},\tau^{\left [ n \right ]}$ to get
the optimal
$\breve{\mathbf{P}}_{c},\big \{ \breve{\mathbf{P}}_{k}\big \}_{k=1}^{K},\breve{\eta},\breve{\beta},\breve{\tau},\breve{\mathrm{objective}}$;
\State $n\leftarrow n+1$; 
\State Update $ \mathbf{P}_{c}^{\left [ n \right ]} \leftarrow \breve{\mathbf{P}}_{c},\big \{ \mathbf{P}_{k}^{\left [ n \right ]} \big \}_{k=1}^{K}\leftarrow \big \{ \breve{\mathbf{P}}_{k}\big \}_{k=1}^{K},\eta^{\left [ n \right ]}\leftarrow \breve{\eta},\beta^{\left [ n \right ]}\leftarrow \breve{\beta},\tau^{\left [ n \right ]}\leftarrow \breve{\tau}, 
\mathrm{objective}^{\left [ n\right ]}\leftarrow \breve{\mathrm{objective}};$
\Until{$\big |\mathrm{objective}^{\left [ n \right ]}-\mathrm{objective}^{\left [ n -1\right ]}  \big |< \varepsilon$ ;}
\end{algorithmic}
\end{algorithm}

\section{Simulation Results}

In this section, simulation results are presented.
We consider a DFRC satellite which is equipped with 
$N_{t}=9$ transmit antennas, and the radar receiver is equipped with $N_{r}=10$ antennas.
The transmit and receive arrays are respectively assumed to be
uniform circular array (UCA) and uniform linear array (ULA).
According to the architecture of single feed per beam (SFPB), 
there are $N_{t}=9$ on board antenna feeds serving $K=9$ beams on the ground. 
Within each beam, uniformly located user terminals are served in a time division multiplexed (TDM) manner \cite{6184256}.
We assume $\sigma _{n,k}^{2}=\sigma _{n}^{2}= 0 \ \mathrm{dBm}, \ \forall k \in \mathcal{K}$.
The total transmit power budget is $P_{t} = 30 \ \mathrm{dBm}$.
Simulation parameters of the introduced satellite channel model are listed in Table \ref{table_example} \cite{9165811, 6184256}.
We assume that the true target parameters are $\theta = 45 \degree,
\phi = 83 \degree, \mathcal{F}_{D} = 2 \ \mathrm{kHz}$,  
and the number of transmit symbols within one coherent processing interval (CPI) is $L= 256$.

\begin{table} 

\caption{Simulation Parameters}
\label{table_example}
\centering
\begin{tabular}{c|c}
\hline
\textbf{Parameter} & \textbf{Value}\\
\hline
Orbit & LEO\\
Carrier frequency &  $20\ \mathrm{GHz}  $\\
Satellite height & $1000\ \mathrm{km}$\\
Bandwidth & 25 MHz\\
3 dB angle & $0.4\degree$ \\
Satellite antenna gain & 17 dBi\\
Boltzmann constant & $1.38 \times 10^{-23} \ \mathrm{J/m}$ \\
System noise temperature & 517 K\\
Rain fading parameters & $\left ( \mu_{\mathrm{rain}},\sigma_{\mathrm{rain}} ^{2} \right )=\left ( -2.6,1.63 \right )$\\
\hline
\end{tabular}
\end{table}


\begin{figure}
\centering
\includegraphics[width=0.8 \columnwidth]{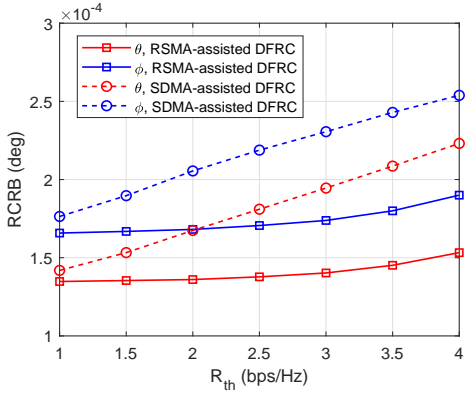}
\caption{The RCRB performance versus $R_{\mathrm{th}}$.
$N_{t} = K=9,\ N_{r} = 10,\ L = 256, \ \mathrm{SNR}_{\mathrm{radar}} = 28\ \mathrm{dB}.$
}
\label{fig:RCRB_RTH}
\end{figure} 

The performance of our proposed RSMA for DFRC beamforming is explicitly shown
in Fig. \ref{fig:RCRB_RTH} in terms of the root-CRB (RCRB).
The benchmark scenario is the conventional SDMA-assisted DFRC beamforming, which can be regarded as a special case of RSMA by turning off the common stream.
For both transmission strategies, we can see the trade-off between radar and communication performance.
As the  communication rate threshold $R_{\mathrm{th}}$ grows, the RCRBs for target estimation become higher.
Moreover, it is observed that the proposed RSMA-assisted DFRC beamforming always outperforms SDMA.
The RCRB of SDMA-assisted DFRC beamforming grows faster as the increment of $R_{\mathrm{th}}$, while the RCRB of RSMA strategy remains at a lower level. 
This implies that SDMA is not robust enough to enable a good trade-off between radar and communication performance, while RSMA is superior to SDMA for the DFRC scenario to manage the inter-beam interference as well as performing the radar functionality.

\begin{figure}
\centering
\includegraphics[width=0.8 \columnwidth]{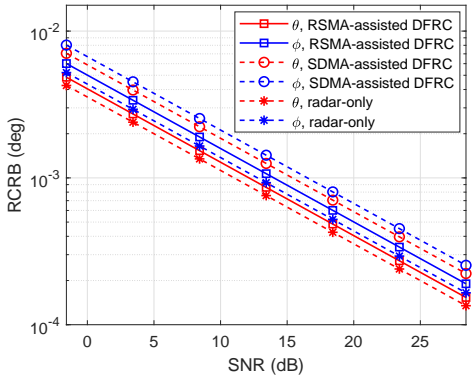}
\caption{The RCRB performance versus  $\mathrm{SNR}_{\mathrm{radar}}$.
$N_{t} = K=9,\ N_{r} = 10,\ L = 256, \ R_{\mathrm{th}} = 4\ \mathrm{bps/Hz}.$
}
\label{fig:RCRB_SNR}
\end{figure}

Fig. \ref{fig:RCRB_SNR} shows the RCRB performance versus radar SNR when $R_{\mathrm{th}} = 4\  \mathrm{bps/Hz}$.
The radar receive SNR is defined as $\mathrm{SNR}_{\mathrm{radar}} = \left | \alpha \right |^{2} P_{t}/\sigma _{m}^{2}$.
We also consider the  radar-only scenario as a benchmark, which can be obtained by removing all communication constraints from the formulated optimization problem.
From Fig. \ref{fig:RCRB_SNR}, the radar-only scenario performs the best as we expected. 
The performance of RSMA-assisted DFRC beamforming is very close the radar-only scenario even if quite strict communication QoS constraints are ensured simultaneously.

Additionally, we analyze the sensing capability at the radar receiver side. 
To extract the Doppler frequency and angle information of the  target from the received radar signal, 
we consider the Capon beamformer, which maximizes the signal-to-noise ratio at the receiver \cite{xu2006radar}
\begin{align}
    \mathbf{w} = \frac{\mathbf{R}_{Z}^{-1} \mathbf{b}\left ( \theta \right )}{ \mathbf{b}\left ( \theta \right )^{H}\mathbf{R}_{Z}^{-1} \mathbf{b}\left ( \theta \right ) },
\end{align}
where $\mathbf{R}_{Z}$ is  the covariance matrix of the received signal
\begin{align}
    \mathbf{R}_{Z} = 
     \frac{1}{L} \sum_{l=1}^{L}  \mathbf{z}\left [ l \right ]  \mathbf{z}\left [ l \right ]^{H}.
\end{align}
The Capon estimate of $\alpha$ is derived by minimizing the following cost-function
\begin{align}
    \widehat{\alpha}  & = \arg\min _{\alpha } \mathbb{E} \left \{ \left | \mathbf{w}^{H} \mathbf{z}\left [ l \right ] - \alpha e^{j 2\pi \mathcal{F}_{D} lT} {\mathbf{a}}^{H} \left ( \theta, \phi \right )\mathbf{x}\left [ l \right ]   \right |^{2} \right \} \notag
    \\
    &=  \mathbb{E} \left \{\frac{e^{-j 2\pi \mathcal{F}_{D} lT} \mathbf{w}^{H}\mathbf{z}\left [ l \right ] \mathbf{x}^{H}\left [ l \right ]    \mathbf{a}\left ( \theta, \phi \right )  }{ \mathbf{a}^{H}\left ( \theta, \phi \right ) \mathbf{x}\left [ l \right ]\mathbf{x}^{H}\left [ l \right ] \mathbf{a}\left ( \theta, \phi \right ) }  \right \}  \notag
    \\ 
    &=
     \frac{\mathbf{w}^{H} \mathbf{Z}  \mathbf{D}^{H} \mathbf{X}^{H}  \mathbf{a}\left ( \theta, \phi \right ) }{L\mathbf{a}^{H}\left ( \theta, \phi \right ) \mathbf{R}_{X}\mathbf{a}\left ( \theta, \phi \right )  },
     \label{capon_estimate}
\end{align}
where $\mathbf{X} = \left [ \mathbf{x}\left [ 1 \right ], \cdots,  \mathbf{x}\left [ L \right ]\right ]$, $\mathbf{Z} = \left [ \mathbf{z}\left [ 1 \right ], \cdots,  \mathbf{z}\left [ L \right ]\right ]$, 
$\mathbf{D} = \mathrm{diag}\left \{  \left [ e^{j 2\pi \mathcal{F}_{D} T}, \cdots, e^{j 2\pi \mathcal{F}_{D} LT} \right ]\right \}$.
Note that $\widehat{\alpha}$ is a function of $\theta,\ \phi,\ \mathcal{F}_{D}$.
Evaluating the expression of
$\widehat{\alpha}$ at each grid point
requires a three-dimensional search, which is computationally expensive.
Therefore, we adopt a two-stage estimation, starting with the Doppler frequency estimation based on the Fast Fourier Transform (FFT) method.
First, we can rewrite the signal at the radar receiver as
\begin{align}
    \mathbf{z}\left [ i \right ] = \alpha  e^{j 2\pi \mathcal{F}_{D} iT_{s}} {\mathbf{b}} \left ( \theta \right )
 {\mathbf{a}}^{H} \left ( \theta, \phi \right )\mathbf{x}\left [ i \right ] + \mathbf{m}\left [ i \right ],
\end{align}
where the sampling period $T_{s}$ is related to the symbol period $T$ through $T =  M_{\mathrm{symb}} T_{s}$. $M_{\mathrm{symb}}$ is some integer, and
the sampling index $i \in \left \{ 1, \cdots, M_{\mathrm{symb}} L \right \}$.
For all $l\in \mathcal{L}$, $\mathfrak{F}_{l}\left ( N_{\mathrm{fft}} \right )$ with grid of size $N_{\mathrm{fft}}$ points, denotes the FFT of the following expression
\begin{align}
    \sum_{n_{r}=1}^{N_{r}} \left [  \mathbf{z}_{n_{r}}\left [ \left ( l-1 \right ) M_{\mathrm{symb}}+1 \right ], \cdots,\mathbf{z}_{n_{r}}\left [ l M_{\mathrm{symb}} \right ] \right ].
\end{align}
The Doppler frequency estimate $\widehat{\mathcal{F}}_{D}$ can be obtained by solving $\arg\max_{n_{\mathrm{fft}}} \big| \sum_{l=1}^{L}\mathfrak{F}_{l}\left ( N_{\mathrm{fft}} \right ) \big | $.
Next, by inserting $\widehat{\mathcal{F}}_{D}$ into (\ref{capon_estimate}), only $\theta$ and $\phi$ remain unknown, and can be estimated using a two-dimensional (2-D) search.

\begin{figure}
\centering
\includegraphics[width=0.8 \columnwidth]{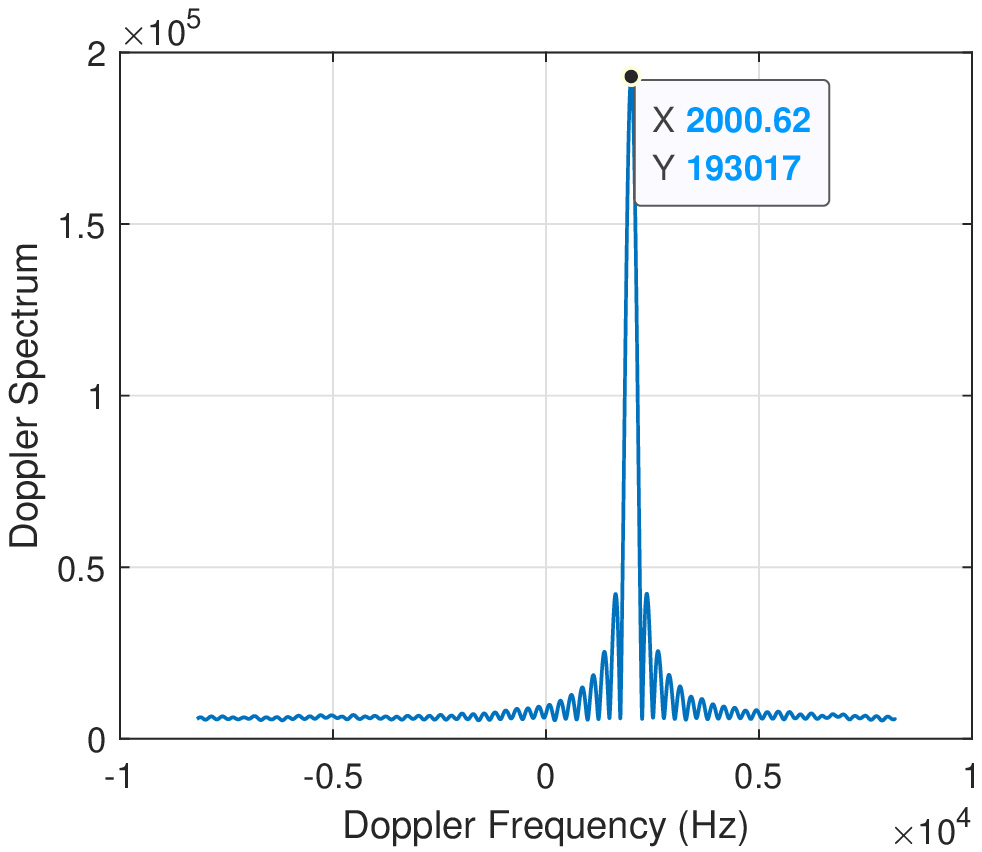}
\caption{ An example of Doppler frequency estimation.
$N_{t} = K=9,\ N_{r} = 10,\ L = 256, \ M_{\mathrm{symb}} = 64, \ R_{\mathrm{th}}=4\ \mathrm{bps/Hz}.$
}
\label{fig:DF}
\end{figure} 

\begin{figure}
\centering
\includegraphics[width=0.85 \columnwidth]{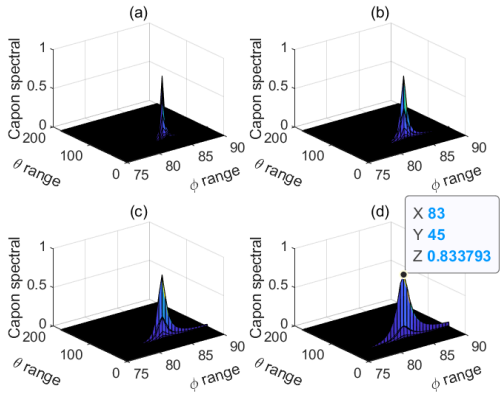}
\caption{ Angle estimation based on the optimized RSMA-assisted DFRC beamforming.
$N_{t} = K=9,\ N_{r} = 10,\ L = 256, \ \mathrm{SNR}_{\mathrm{radar}} = 28\ \mathrm{dB}.$
(a) $R_{\mathrm{th}}=1\ \mathrm{bps/Hz}$, (b) $R_{\mathrm{th}}=2\ \mathrm{bps/Hz}$, (c) $R_{\mathrm{th}}=3\ \mathrm{bps/Hz}$, (d) $R_{\mathrm{th}}=4\ \mathrm{bps/Hz}$.
}
\label{fig:CAPON_RTH}
\end{figure}

\begin{figure}
\centering
\includegraphics[width=0.85 \columnwidth]{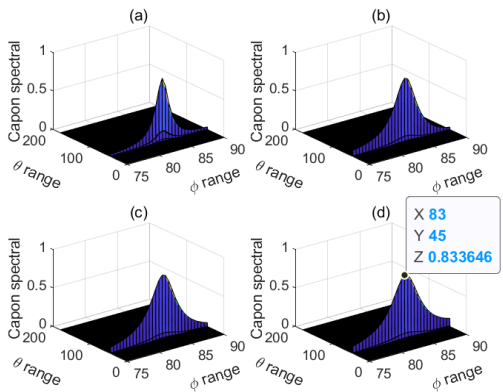}
\caption{ Angle estimation based on the optimized SDMA-assisted DFRC beamforming.
$N_{t} = K=9,\ N_{r} = 10,\ L = 256, \ \mathrm{SNR}_{\mathrm{radar}} = 28\ \mathrm{dB}.$
(a) $R_{\mathrm{th}}=1\ \mathrm{bps/Hz}$, (b) $R_{\mathrm{th}}=2\ \mathrm{bps/Hz}$, (c) $R_{\mathrm{th}}=3\ \mathrm{bps/Hz}$, (d) $R_{\mathrm{th}}=4\ \mathrm{bps/Hz}$.
}
\label{fig:CAPON_RTH_sdma}
\end{figure}

Throughout simulation, the transmitted DFRC signals are beamformed uncorrelated QPSK symbols as given in (\ref{transmit}).
Beamforming matrices are the optimized ones obtained by solving the above formulated problem.
Fig. \ref{fig:DF} shows an example of the Doppler frequency estimation.
Fig. \ref{fig:CAPON_RTH} and 
Fig. \ref{fig:CAPON_RTH_sdma} respectively show the 2-D angle estimation of RSMA- and SDMA-assisted DFRC beamforming with different $R_{\mathrm{th}}$ settings.
We can observe that 
RSMA  achieves sharper  peaks and thus better estimation performance compared with SDMA.
For both scenarios, the performance decreases as  $R_{\mathrm{th}}$ becomes higher, which matches the theoretical RCRB-$R_{\mathrm{th}}$ trade-off shown in Fig. \ref{fig:RCRB_RTH}.

\section{Conclusion}
We have investigated the application of RSMA to a multi-antenna bistatic DFRC satellite system,
which simultaneously serves multiple communication users and detects
a moving target.
The DFRC beamforming is designed by formulating optimization problems to minimize the CRB of target estimation while guaranteeing
the QoS constraints of SUs.
An iterative algorithm based on SCA is proposed to solve the problem.
Based on the optimized DFRC waveform and the Capon method, parameter estimation of
the moving target is evaluated at the radar receiver side. Through simulation results, RSMA is demonstrated to be very promising for the DFRC satellite system to enable a better communication-sensing trade-off and achieve better
target estimation performance than the benchmark strategy.

\bibliographystyle{IEEEtran}
\bibliography{ref}

\end{document}